\documentclass[DIV=20,10pt,a5paper,onecolumn,headings=normal]{scrartcl}

\usepackage{indentfirst}
\usepackage[english]{babel}
\usepackage{graphicx}				
\usepackage{amsmath}
\usepackage{amssymb}
\usepackage{bm}	
\usepackage[title]{appendix} 
\usepackage{cite} 
\usepackage[unicode, pdfstartview= FitH]{hyperref} 
\usepackage{scrpage2} 
\usepackage{geometry} 
\def\({\small(\normalsize\setlength{\abovedisplayskip}{6pt plus 1pt minus 1pt}\setlength{\belowdisplayskip}{4pt plus 1pt minus 1pt}} 
\def\){\small)\normalsize\setlength{\abovedisplayskip}{6pt plus 1pt minus 1pt}\setlength{\belowdisplayskip}{4pt plus 1pt minus 1pt}}

\clearscrheadfoot	

\begin{document}

\setlength{\abovedisplayskip}{6pt plus 1pt minus 1pt}  
\setlength{\belowdisplayskip}{4pt plus 1pt minus 1pt}  

\renewcommand{\abstractname}{\vspace{-50pt}}

\title{Statistical mechanics of fluids in a step potential}

\author{V.M. Zaskulnikov \thanks{zaskulnikov@gmail.com}}
\date{\today}

\maketitle

\begin{abstract}

The paper deals with the problem of surface effects at a fluid boundary produced by a step force field. A classical simple fluid with a locally placed field simulating a solid is considered.
 
The specific surface $\Omega$-potential $\gamma$, the surface number density, and the Henry adsorption constant are determined.
 
A surface cluster expansion (the expansion of the surface omega potential in powers of the activity) is obtained. This expansion is similar to the cluster expansion for the pressure in which the integrals of the Ursell factors are replaced by sums of the first-order moments of the Ursell factors over sectors or orthants.
 
The contact theorem is extended to the case of a finite step field.
 
It is found that the surface number density (its invariant part) is determined by the first-order moment of the pair Ursell function taken over the entire space.
 
The high-and low-temperature limits are analyzed and are shown to be consistent with the previously obtained general results.
 
Symmetry features of the solution with respect to permutation of the solid and fluid regions are established.

\vspace{20pt}

\end{abstract}

\section{\label{sec:01}Introduction}

Statistical mechanics of fluids near an arbitrary permeable wall was constructed in \cite{Zaskulnikov201111a}. In the present paper, this theory is developed and applied to the adsorption problem with a finite step potential. In this case, an important role is played by the specially designed apparatus of sectorial and orthant moments.
 
The importance of this problem is due to two things. First, the step potential is an important model system. Suffice it to say that this problem is an extension of the problem of a rigid impermeable wall and, in particular, it provides a generalization of the long and widely used contact theorem \cite{Percus1976} \(section \ref{sec:09}\).
 
In this model system, surface cluster expansion plays an important role, primarily as an accurate method for validating any results in the range of not too high number densities \(sections \ref{sec:05}, \ref{sec:06}, and appendix \ref{sec:b}\).
 
Second, the step potential model is related to probabilistic problems. In particular, the behavior of the nonlinear surface coefficient at high temperatures is related to the surface part of the standard deviation of the number density \(section \ref{sec:07}\).
 
However, as noted in \cite{SteckiSokolowski1980}, even the general problem of a permeable wall had not been considered in terms of statistical mechanics until 1980. Later, the emphasis has shifted to approximate and computational methods.
 
A number of results valid for the step potential is not considered in the present paper because they are formulated in general terms in \cite{Zaskulnikov201111a}. These are, for example, a proof of the identity of the ``tension'' and ``adsorption'' approaches, a consideration of the ``mechanical definition'' of $\gamma$, and other issues.

\section{\label{sec:02}Primary relations}

\subsection{\label{subsec:021}Canonical ensemble}

The probability density of finding a given spatial configuration of a particular set of particles \cite[p.\,181]{hillstatmeh1956} is given by
\begin{equation}
P^{(k)}_{1...k} = \frac{1}{Z_N}\int \limits_V \exp(-\beta U_{1...N})d\bm{r}_{k+1}...d\bm{r}_N. 
\label{eq:001}
\end{equation}

Here $N$ is the number of particles in the system, $\beta = 1/k_BT$, $k_B$ is the Boltzmann constant, $T$ is the temperature, $U_{1 ... N}$ is the particle interaction energy, and $V$ the volume of the system. The integration is performed over the coordinates of the particles of the ensemble. $Z_N$ is the configuration integral
\begin{equation}
Z_N = \int \limits_V \exp(-\beta U_{1...N}) d\bm{r}_1...d\bm{r}_N 
\label{eq:002}.
\end{equation}

Passing to the distribution functions for an arbitrary set of particles, we obtain
\begin{equation}
\varrho^{(k)}_{C,1...k} = \frac{N!}{(N-k)!}P^{(k)}_{1...k} 
\label{eq:003}.
\end{equation}

This quantity defines the probability density of finding a given configuration of $k$ arbitrary particles for a canonical ensemble.

\subsection{\label{subsec:022}Grand canonical ensemble (GCE)}

We average the equality \(\ref{eq:003}\) over fluctuations in the number of particles, i.\,e., apply the operation $\sum_{N=0}^\infty P_N^V$ to both sides of the equation. Here
\begin{equation}
P_N^V = \frac{z^N Z_N}{N!\Xi_V} 
\label{eq:004}
\end{equation}
is the probability for the GCE to have a certain number of particles $N$ within volume $V$. $z$ is the activity:
\begin{equation}
z = \frac{e^{\mu/k_BT}}{\Lambda^3} 
\label{eq:005},
\end{equation}
where $\mu$ is the chemical potential and $\Lambda = h/\sqrt[]{2 \pi mk_BT}$; $h$ is the Planck constant, $m$ is the particle mass, and $\Xi_V$ is the grand partition function of a system of volume $V$:
\begin{equation}
\Xi_V = 1 + \sum_{N=1}^\infty \frac{z^N Z_N}{N!} 
\label{eq:006}.
\end{equation}

We obtain:
\begin{equation}
\varrho^{(k)}_{G,1...k} = \sum_{N=k}^\infty \varrho^{(k)}_{C,1...k} P_N^V
\label{eq:007},
\end{equation}
or
\begin{equation}
\varrho^{(k)}_{G,1...k} = \frac{z^k}{\Xi_V} \Big [ \exp{(-\beta U_{1...k})} +  \sum_{N=1}^\infty \frac{z^N}{N!} \int \limits_V    \exp(-\beta U_{1...N+k})d\bm{r}_{k+1}...d\bm{r}_{k+N}\Big ]
\label{eq:008}.
\end{equation}

$\varrho^{(k)}_{G,1...k}$ define the probability density of finding a certain configuration of $k$ arbitrary particles in the GCE. For an ideal gas, this quantity is equal to $\varrho^{k}$, where $\varrho = \overline{N}/V$ is the number density.

\subsection{\label{subsec:023}Ursell factors}

The Ursell factors ${\cal U}^{(k)}_{1...k}$ are also called cluster functions. These functions appear in the well-known pressure expansion in powers of the activity \cite [p.\,135]{hillstatmeh1956}\footnote{In this paper, the expansion is not in integrals of the Ursell factors, but in their related Thiele semi-invariants. The relationship between them is defined by the transition from particular partitions to a partition topology \scriptsize(\footnotesize see below\scriptsize)\footnotesize.}, \cite [p.\,230]{landaulifshitz1980}
\begin{equation}
P(z,T) =zk_BT  + k_BT \sum_{n=2}^\infty \frac{z^n}{n!} \int {\cal U}^{(n)}_{1...n} d\bm{r}_2...d\bm{r}_n.
\label{eq:009}\footnote{The absence of the integration limits will always mean that the integration is over infinite space. The validity of infinite limits for this case is shown in \cite{zaskulnikov200911a}.}
\end{equation}

They are also included in the corresponding expansion of the number density
\begin{equation}
\varrho(z) = z + z\sum_{n=1}^\infty \frac{z^n}{n!} \int {\cal U}^{(n+1)}_{1...n+1} d\bm{r}_{2}...d\bm{r}_{n+1}
\label{eq:010},
\end{equation}
which is obvious from the formula
\begin{equation}
\varrho = \bigg ( \frac{\partial P}{\partial \mu} \bigg )_{ T }
\label{eq:011},
\end{equation}
taking into account the relationship (\ref{eq:005}).

The Ursell factors have the locality property, which is important for our consideration: they decay rapidly as any group of particles, including unit groups, move away from the others.

The Ursell factors can be defined by the equality \cite{ursell1927}
\begin{equation}
\begin{gathered}
{\cal U}^{(k)}_{1...k} = \sum_{\{\bm{n}\}}(-1)^{l-1}(l-1)!\prod_{\alpha = 1}^l \exp[-\beta U(\bm{n}_\alpha)]   \\
1  \leq  ~ k_\alpha \leq k, ~~~ \sum_{\alpha = 1}^l k_\alpha = k, ~~~ \exp(-\beta U_{i}) = 1,  
\label{eq:012}
\end{gathered}
\end{equation}
where $\{\bm{n}\} $ denotes some partition of a given set of $k$ particles with coordinates $\bm{r}_1,...\bm{r}_k$ into disjoint groups $\bm{n}_\alpha$, $l$ is the number of groups in the particular partition, $k_\alpha$ is the size of the group with the number $\alpha $, and $U(\bm{n}_\alpha)$ is the particle interaction energy in the group $\bm{n}_\alpha$. The summation is over all possible partitions, and the meaning of the condition $\exp (- \beta U_ {i}) = 1 $ is obvious: unit groups do not contribute to the product in this case.

For example,
\setlength{\lineskiplimit}{-2pt}
\begin{equation}
\begin{aligned}
&{\cal U}^{(1)}_{1}\mkern 9mu = 1 \\
&{\cal U}^{(2)}_{1,2} \mkern 8mu = \exp(-\beta U_{1,2}) - 1\\
&{\cal U}^{(3)}_{1,2,3} = \exp(-\beta U_{1,2,3}) - \exp(-\beta U_{1,2})- \exp(-\beta U_{1,3}) - \exp(-\beta U_{2,3}) + 2.
\end{aligned}
\label{eq:013}
\end{equation}
\setlength{\lineskiplimit}{-5pt}

\subsection{\label{subsec:024}Presence of an external field}

The configuration integral of a closed inhomogeneous system is given by
\begin{equation}
Z^U_N = \int\limits_{V} \exp(-\beta\sum_{i=1}^N u_i-\beta U_{1...N}) d\bm{r}_1...d\bm{r}_N, \\
\label{eq:014}
\end{equation}
where $u_i \equiv u(\bm{r_i}) $ is the energy of interaction of the $i$-th particle with the field.
 
For the GCE, we introduce the quantity
\begin{equation}
\Xi^U_V = 1+ \sum_{N=1}^\infty \frac{z^N Z^U_N}{N!}, 
\label{eq:015}
\end{equation}
which is obviously the grand partition function of the system in the presence of an external field.
 
For arbitrary activities, the distribution functions have the form
\begin{multline}
\varrho^{(k)}_{G,1\dots k} = \frac{z^k}{\Xi_V^U} \Big [ \prod_{i = 1}^{k} \theta_i \Big ] \Big \{ \exp\big( - \beta U_{1...k}\big)  \\ +  \sum_{N=1}^\infty \frac{z^N}{N!} \int \limits_{V}  \Big [ \prod_{i = k+1}^{k+N} \theta_i \Big ]  \exp\big( - \beta U_{1...N+k}\big)d\bm{r}_{k+1}...d\bm{r}_{N+k} \Big \}
\label{eq:016},
\end{multline}
where
\begin{equation}
\theta_i =  \exp(-\beta u_i)
\label{eq:017}.
\end{equation}

Varying the external potential, we obtain
\begin{equation}
\delta\varrho^{(k)}_{1\dots k} = -\beta \Big [\varrho^{(k)}_{1\dots k}  \sum_{i=1}^k \delta u_i + \int \big ( \varrho^{(k+1)}_{1\dots k+1} - \varrho^{(k)}_{1\dots k} \varrho^{(1)}_{k+1}\big ) \delta u_{k+1} d\bm{r}_{k+1} \Big ]
\label{eq:018},
\end{equation}
which for the case $k=1$ reduces to the Yvon equation
\begin{equation}
\delta\varrho_1 = -\beta \big (\varrho_1\delta u_1 + \int \mathcal{F}^{(2)}_{1,2} \delta u_2 d\bm{r}_2 \big )
\label{eq:019},
\end{equation}
where
\begin{equation}
{\cal F}^{(2)}_{1,2}  =  \varrho^{(2)}_{1,2} -  \varrho_1\varrho_2
\label{eq:020}
\end{equation}
is a second-order Ursell function \(in this case, for the system in the field \).

If the variation of the potential is constant in space\footnote{It is sufficient that this region includes all the coordinates of the initial distribution function and that to their distance from the boundary of this region to the coordinates is at least a few atomic layers. The potential in this case may be arbitrarily inhomogeneous.}, it is equal to the change in the chemical potential with the opposite sign, and from \(\ref{eq:018}\) we obtain
\begin{equation}
\int \big ( \varrho^{(k+1)}_{1\dots k+1} - \varrho^{(k)}_{1\dots k} \varrho^{(1)}_{k+1}\big ) d\bm{r}_{k+1} = z^{k+1} \frac{\partial}{\partial z}\bigg ( \frac{\varrho^{(k)}_{1\dots k}}{z^k} \bigg )
\label{eq:021},
\end{equation}
where the differentiation is performed at constant temperature. Thus, this relation has the same form as in the case of a homogeneous medium \cite{zaskulnikov200911a}.

Using the properties of the fractional generating function \(\ref{eq:008}\) \(see \cite{Zaskulnikov201111a}\), we obtain an analog of \(\ref{eq:010}\) 
\begin{equation}
\varrho(\bm{r_1},z) = z \theta_1 + z\theta_1\sum_{n=1}^\infty \frac{z^n}{n!} \int \Big [ \prod_{i = 2}^{n+1} \theta_i \Big ]{\cal U}^{(n+1)}_{1...n+1} d\bm{r}_{2}...d\bm{r}_{n+1}
\label{eq:022},
\end{equation}
where $\varrho(\bm{r_1},z)$ is the number density in the presence of an external field.
 
We will also need the following expansion:
\begin{equation}
{\cal F}^{(2)}_{12}(z)= z^2\theta_1\theta_2 \Big \{ {\cal  U}^{(2)}_{12} + \sum_{n=1}^\infty \frac{z^n}{n!}\int \Big [ \prod_{l = 3}^{n+2} \theta_l \Big ] {\cal  U}^{(n+2)}_{1...n+2} d\bm{r}_{3}...d\bm{r}_{n+2} \Big \}
\label{eq:023}.
\end{equation}

An analog of \(\ref{eq:023}\) for a homogeneous medium is obtained in \cite{UhlenbeckFord1962}. For an inhomogeneous medium, it is sufficient to note that this relation is provided by the same recurrence relations for the partial localization factors \cite{Zaskulnikov201111a} as in the homogeneous case.

\section{\label{sec:03}General case}

This section briefly summarizes the results obtained in \cite{Zaskulnikov201111a} for the case of a permeable wall with the near-surface potential of arbitrary form.

It is shown that
\begin{equation}
\Omega^U = -P(z)V +\big[P(z)-P(z')\big]\int \widetilde{\varphi}(\bm{r}) d\bm{r} + \nu(z, z') A,
\label{eq:024}
\end{equation}
where $\Omega^U$ is the omega potential  of the system with the field placed in it, $A$ is the bounding area of the field effect region, and
\begin{equation}
\nu (z,z') =  \int\limits_{-\infty}^{\infty}\Big [ \widetilde{\theta}(x) P(z)+\widetilde{\varphi}(x)P(z')- P^*(x,z,z')  \Big ] dx
\label{eq:025}
\end{equation}
is a nonlinear surface coefficient \($x$ is the coordinate directed along the field gradient\).
 
Here we have introduced the local pressure
\begin{equation}
P^{*}(\bm{r}) =  \int\limits_{-\infty}^{\mu} \varrho(\bm{r}) d\mu',
\label{eq:026}  
\end{equation}
which, in this case, is equivalent to
\begin{equation}
\varrho(\bm{r}) = \bigg (\frac{\partial P^{*}(\bm{r})}{\partial \mu} \bigg )_{T,u}    
\label{eq:027}.
\end{equation}

In addition,
\begin{equation}
z' = z \exp {(- \beta u_0)}
\label{eq:028}
\end{equation}
is the internal activity in the field region, $u_0$ is the potential of the external field in the depth of the field effect region, and the quantities
\begin{equation}
\widetilde{\theta}(x) = \frac{\exp {[- \beta u(x)]} - \exp {(- \beta u_0)}}{1 - \exp {(- \beta u_0)}}
\label{eq:029}
\end{equation}
and
\begin{equation}
\widetilde{\varphi}(x) = \frac{1 - \exp {[- \beta u(x)]} }{1 - \exp {(- \beta u_0)}}, 
\label{eq:030}
\end{equation}
generalize the Boltzmann factors \(\ref{eq:017}\) and the Mayer functions
\begin{equation}
\varphi_i = 1 - \exp {(- \beta u_i)} 
\label{eq:031}
\end{equation} 
respectively. It is easy to see that
\begin{equation}
\widetilde{\theta}_i + \widetilde{\varphi}_i = 1
\label{eq:032}
\end{equation}
and that the functions $\widetilde {\theta} _i $ and $\widetilde {\varphi} _i $ decay in and out of the field, respectively, and tend to 1 otherwise.

In addition, they have the property
\begin{equation}
z\widetilde{\theta}_i + z'\widetilde{\varphi}_i = z\theta_i
\label{eq:033}.
\end{equation}

The nonlinear surface coefficient is expanded in a series in the activity as
\begin{equation}
\nu (z,z') =  k_BT \sum_{n=2}^\infty \frac{1}{n!}\int {g}^{(n-1)}_{2...n}  {\cal U}^{(n)}_{0,2...n} d\bm{r}_2...d\bm{r}_k,
\label{eq:034}
\end{equation}
where
\begin{equation}
{g}^{(n-1)}_{2...n}= \int \limits_{- \infty}^{+ \infty}   \Big \{\widetilde{\theta}(x) z^n +\widetilde{\varphi}(x) z'^n- \big [z \widetilde{\theta}(x) + z' \widetilde{\varphi}(x)\big ]\prod_{i = 2}^n \big [z \widetilde{\theta}(x+x_i) + z' \widetilde{\varphi}(x+x_i) \big] \Big \} d x
\label{eq:035}
\end{equation}
is a function of the variables $x_2, \dots, x_n$ which is symmetric under permutations.

The expression \(\ref{eq:024}\) can be transformed to
\begin{equation}
\Omega = -P(z)(V-V') - P(z')V'  + \gamma(x')  A,
\label{eq:036}
\end{equation}
where
\begin{equation}
\gamma(x') = - \big [P(z)-P(z') \big ](x'-x_0) + \nu (z,z') 
\label{eq:037},
\end{equation}
$x'$ is some arbitrary point within the transition zone or near it which defines the volume $V'$, and
\begin{equation}
x_0 = - \int\limits_{-\infty}^{0} \widetilde{\theta}(x) dx + \int\limits_{0}^{\infty} \widetilde{\varphi}(x) dx
\label{eq:038}
\end{equation}
defines the zero adsorption plane \(see below\).

The expression \(\ref{eq:037}\) gives the general form of the specific surface $\Omega$-potential. The last expression is divided into a term linear in pressure and the nonlinear term $\nu $.
 
We see that there is a certain degree of arbitrariness in the division into volume and surface terms. However, for the step potential, its role is not as important, due to the presence of a distinct boundary.
 
Another form equivalent to \(\ref{eq:037}\) is
\begin{equation}
\gamma(x') =  \int\limits_{-\infty}^{x'} \big [P(z') - P^{*}(x)\big ] dx +  \int\limits_{x'}^{+\infty} \big [P(z) - P^{*}(x)\big ] dx
\label{eq:039}. 
\end{equation}

Differentiating \(\ref{eq:037}\) with respect to the chemical potential and taking into account that the relation
\begin{equation}
\varrho_s =  -\bigg ( \frac{\partial \gamma}{\partial \mu} \bigg )_{ T }
\label{eq:040}
\end{equation}
retains its form, we obtain another expression for the surface number density for an arbitrary position of the separating surface:
\begin{equation}
\varrho_s =  \big [\varrho(z) - \varrho(z')\big ] (x'-x_0)+ \int\limits_{-\infty}^{+\infty} \big [ \varrho(\bm{r}) - \widetilde{\theta}(x) \varrho(z) - \widetilde{\varphi}(x) \varrho(z') \big ]  dx
\label{eq:041}.
\end{equation}
Here the parts linear and nonlinear in number density are clearly separated. When differentiating $\nu(z,z')$ \(and other quantities that depend on $z'$\) with respect to $z$, the quantity $z'$ must obviously be considered a function of $z$, in accordance with \(\ref{eq:028}\).

Obviously,
\begin{equation}
\varrho_s = \varrho_{s,l} + \varrho_{s,n}
\label{eq:042},
\end{equation}
where the linear part is defined by the equation
\begin{equation}
\varrho_{s,l} = \big[\varrho(z) - \varrho(z')\big] (x'-x_0),
\label{eq:043}
\end{equation}
and the nonlinear part by
\begin{equation}
\varrho_{s,n} = \int\limits_{-\infty}^{+\infty}\big [ \varrho(\bm{r}) - \widetilde{\theta}(x) \varrho(z) - \widetilde{\varphi}(x) \varrho(z') \big]  dx
\label{eq:044}.
\end{equation}

In addition, it is obvious that
\begin{equation}
\varrho_{s,l} =   (x'-x_0)\bigg ( \frac{\partial  \big [P(z)-P(z')\big]}{\partial \mu} \bigg )_{ T }
\label{eq:045}
\end{equation}
and
\begin{equation}
\varrho_{s,n} =  - \bigg ( \frac{\partial \nu(z,z')}{\partial \mu} \bigg )_{ T }
\label{eq:046}.
\end{equation}

For the Henry adsorption constant
\begin{equation}
K_H = \lim_{\varrho\rightarrow 0} \frac{\varrho_s}{\varrho}
\label{eq:047}  
\end{equation}
we have  
\begin{equation}
K_H(x') = \big[x' - x_0(T)\big]\big[1 - \exp(-\beta u_0)\big]
\label{eq:048}
\end{equation}
and the problem of determining $K_H$ reduces to the calculation of $x_0$.

Note that $K_H (x_0) = 0$, and the surface through $x_0$ can therefore be called the surface of zero adsorption.
 
Using the expansion \(\ref{eq:022}\), for the Henry absorption constant, we obviously have
\begin{equation}
k_H = \lim_{\varrho\rightarrow 0} \frac{\varrho(z')}{\varrho(z)} = \exp(-\beta u_0)
\label{eq:049}.
\end{equation}

\section{\label{sec:04}Basic equations}

We now consider the step potential.

We examine an open statistical system in a field such that the field effect area is smaller than the size of the system itself and the field is away from its boundaries; i.\,e., this is still the case of ``a field in a system''.

The potential of the external fields is assumed to have a stepped form
\begin{equation}
u(x)  = 
	 \left\{ 
			\begin{array}{ll} 
         u_0 & (x<x_{step})\\   
         0 & (x>x_{step}),
     	\end{array}  
		\right.
			\label{eq:050}
\end{equation}
where $x_{step}$ is the coordinate at which the potential energy of the particles has a step discontinuity, and the interparticle potential is arbitrary, provided that it decays rapidly enough for the convergence of the zero- and first-order moments of the Ursell factors.

We introduce the characteristic functions
\begin{equation}
\psi(x) = 
	 \left\{ 
			\begin{array}{ll} 
         1 & (x<0)\\   
         0 & (x>0)
     	\end{array}  
		\right.
			\label{eq:051}
\end{equation}
\begin{equation}
\chi(x) = 
	 \left\{ 
			\begin{array}{ll} 
         0 & (x<0)\\   
         1 & (x>0),
     	\end{array}  
		\right.
			\label{eq:052}
\end{equation}

Then, for the step potential, we have
\begin{equation}
\widetilde{\theta}(x) = \chi(x-x_{step})  
\label{eq:053}
\end{equation}
\begin{equation}
\widetilde{\varphi}(x) = \psi(x-x_{step})  
\label{eq:054}
\end{equation}
and from \(\ref{eq:025}\) we obtain
\begin{equation}
\nu(z,z') =  \int\limits_{-\infty}^{x_{step}} \big[P(z') - P^{*}(x,z,z')\big] dx +  \int\limits_{x_{step}}^{+\infty}\big [P(z) - P^{*}(x,z,z')\big] dx
\label{eq:055}.  
\end{equation}

From \(\ref{eq:038}\), it is easy to find for this case that
\begin{equation}
x_0 = x_{step}
\label{eq:056},
\end{equation}
and \(\ref{eq:037}\) becomes view
\begin{equation}
\gamma(x',z,z') = - \big[P(z)-P(z')\big](x'-x_{step}) + \nu (z,z') 
\label{eq:057}.
\end{equation}

The form of \(\ref{eq:039}\), of course, remains unchanged.

As in the general case \cite{Zaskulnikov201111a}, for the step potential, we can make the substitutions $P^* \rightarrow P_t$ and $P^* \rightarrow P_{st}$ in the basic equations \(\ref{eq:055}\) and \(\ref{eq:039}\), where $P_t $ is the tangential component of the pressure tensor  and $P_{st} $ is the pressure on the transverse wall.
 
For the surface number density, from \(\ref{eq:041}\) we obtain
\begin{equation}
\varrho_s =  \big[\varrho(z) - \varrho(z')\big] (x'-x_{step})+ \int\limits_{-\infty}^{x_{step}}\big [ \varrho(x) -  \varrho(z') \big ]  dx + \int\limits_{x_{step}}^{\infty}\big [ \varrho(x) -  \varrho(z) \big ]  dx
\label{eq:058}.
\end{equation}

In the case of the step potential considered in this paper, it is reasonable to choose the separating boundary $x'$ to be equal to $x_{step}$ or $x_{step}\pm D/2$, where $D$ is a parameter close to the particle diameter. The last variant is convenient because it includes the ``dead'' \(conventionally speaking\) volume, allowing the volume of one of the phases to be conserved.
 
For the Henry adsorption constant, from \(\ref{eq:048}\) we have
\begin{equation}
K_H(x') = (x' - x_{step})\big[1 - \exp(-\beta u_0)\big]
\label{eq:059}.
\end{equation}

The meaning of the relation \(\ref{eq:059}\) is obvious: it is a coefficient proportional to the difference between the number densities in the limit of low activities $[1 - \exp(-\beta u_0)]$ multiplied by the volume of a parallelepiped of height $[x' - x_{step}]$ and unit area at the base.
 
In this case, using \(\ref{eq:049}\), we arrive at the relation
\begin{equation}
K_H(x') = (x' - x_{step})(1 - k_H)
\label{eq:060},
\end{equation}
between the  Henry adsorption and absorption constants. Generally speaking, the relationship between these two constants is parametric - through temperature \cite{Zaskulnikov201111a}, but in this case, it can be considered conventional.
 
As in the general case of positive potentials \cite{Zaskulnikov201111a}, from \(\ref{eq:060}\) it is evident that the adsorption and absorption in the range $0 < k_H < 1$ are antagonistic. For a given $x '$, the adsorption is maximal at minimal absorption and tends to zero as $k_H \rightarrow 1$ \(100\% solubility\).
 
For $k_H> 1$ \(negative potentials\), the problem is redefined by symmetry \(section \ref{sec:08}\) and again reduces to the interval considered.

\section{\label{sec:05}Surface cluster expansion}

We calculate the nonlinear surface coefficient using the first-order orthant moments technique \(appendix \ref{subsec:a03}\). Our goal is to simplify the series \(\ref{eq:034}\), or reduce it to a known quantity \(this will be done in the next section\).
 
In \(\ref{eq:034}\), passing to the function ${f}^{(n-1)} = {g}^{(n-1)}/z^n$, we have
\begin{equation}
\nu (z,z') =  k_BT \sum_{n=2}^\infty \frac{z^n}{n!}\int {f}^{(n-1)}_{2...n}  {\cal U}^{(n)}_{0,2...n} d\bm{r}_2...d\bm{r}_n,
\label{eq:061}
\end{equation}
where
\begin{equation}
{f}^{(n-1)}_{2...n}= \int \limits_{- \infty}^{+ \infty}   \Big \{\widetilde{\theta}(x)  +\widetilde{\varphi}(x) \lambda^n -\big[ \widetilde{\theta}(x) + \lambda \widetilde{\varphi}(x)\big]\prod_{i = 2}^n \big[ \widetilde{\theta}(x+x_i) + \lambda \widetilde{\varphi}(x+x_i)\big] \Big \} d x
\label{eq:062}.
\end{equation}

Here 
\begin{equation}
\lambda = \frac{z'}{z} = e^{-\beta u_0}
\label{eq:063}.
\end{equation}

Eliminating $\widetilde{\varphi}$ in view of \(\ref{eq:032}\) and using \(\ref{eq:053}\), we have
\begin{equation}
{f}^{(n-1)}_{2...n} = \int \limits_{- \infty}^{+ \infty}   \Big \{\lambda^n +(1-\lambda^n)\chi(x) -\big[\lambda +(1-\lambda) \chi(x)\big]\prod_{i = 2}^n \big[ \lambda +(1-\lambda)\chi(x+x_i)\big] \Big \} d x
\label{eq:064}.
\end{equation}

We expand the product and add and subtract $\chi(x)$ with the corresponding factor to each term, taking into account that 
\begin{equation}
\int \limits_{- \infty}^{+ \infty}   \Big [ \prod_{i = 1}^k  \chi(x+x_i)-\chi(x)     \Big ] d x = \min (x_1, \dots, x_k )
\label{eq:065}.
\end{equation}

The total coefficient of $\chi (x) $ vanishes and we obtain
\begin{equation}
{f}^{(n-1)}_{2...n} = - \sum_{i=1}^n \lambda^{n-i}(1-\lambda)^i \sum_{samp} \min (x_{\alpha_1}, x_{\alpha_2}, \dots, x_{\alpha_i})
\label{eq:066}.
\end{equation}

The inner sum is taken over all subgroups of size $i$ of the total group of $n$ elements \(zero plays the role of the first element\).
 
The Ursell factors are invariant under coordinate inversion\footnote{We use the quantum mechanical definition of the inversion \cite [p.\,96]{landaulifshitz1977}}
\begin{equation}
\bm{r}'_i = - \bm{r}_i
\label{eq:067},
\end{equation}
where $i = 1,\dots, n$, which implies that for an arbitrary configuration,
\begin{equation}
{\cal U}^{(n)}(\bm{r}_{i_1},\bm{r}_{i_2},\dots,\bm{r}_{i_n}) = {\cal U}^{(n)}(-\bm{r}_{i_1},-\bm{r}_{i_2},\dots,-\bm{r}_{i_n}) 
\label{eq:068}.
\end{equation}

Thus, we can equally well assume that
\begin{equation}
{f}^{(n-1)}_{2...n} =  \sum_{i=1}^n \lambda^{n-i}(1-\lambda)^i \sum_{samp} \max (x_{\alpha_1}, x_{\alpha_2}, \dots, x_{\alpha_i})
\label{eq:069}.
\end{equation}

Denoting
\begin{equation}
I_n =  \int {f}^{(n-1)}_{2...n}  {\cal U}^{(n)}_{0,2...n} d\bm{r}_2...d\bm{r}_n,
\label{eq:070}
\end{equation}
from \(\ref{eq:061}\), we have
\begin{equation}
\beta\nu (z,\lambda) =   \sum_{n=2}^\infty \frac{z^n}{n!}I_n
\label{eq:071}.
\end{equation}

From \(\ref{eq:069}\), we obtain
\begin{multline}
I_n =  \sum_{i=2}^{n-1} \lambda^{n-i}(1-\lambda)^i \Big [ \binom{n-1}{i-1}(i-1) \int_{x_2> 0, x_3, \dots, x_i} x_2  {\cal U}^{(n)}_{0,2...n} d\bm{r}_2...d\bm{r}_n  \\ +  \binom{n-1}{i}i \int_{x_2> x_3, \dots, x_{i+1}} x_2  {\cal U}^{(n)}_{0,2...n} d\bm{r}_2...d\bm{r}_n  \Big ] \\ + (1-\lambda)^n (n-1)\int_{x_2> 0, x_3, \dots, x_n} x_2  {\cal U}^{(n)}_{0,2...n} d\bm{r}_2...d\bm{r}_n 
\label{eq:072},
\end{multline} 
where the condition
\begin{equation}
x_{\alpha_1}> x_{\alpha_2}, \dots, x_{\alpha_j}
\label{eq:073}
\end{equation}
implies the simultaneous satisfaction of the inequalities
\begin{equation}
x_{\alpha_1}> x_{\alpha_2}, \dots,		x_{\alpha_1}> x_{\alpha_j}
\label{eq:074}.
\end{equation}

In \(\ref{eq:072}\), we used the fact that all groups of equal size give the same result and that unit groups vanish by antisymmetry. The presence or absence of zero in a group leads to terms of different forms, and the group of the maximum size always contains zero.
 
Consider the first group of terms on the right-hand side of \(\ref{eq:072}\). Making the change of variables
\begin{equation}
\bm{r}'_i =  
	 \left\{ 
			\begin{array}{ll} 
         \bm{r}_{2} & (i = 2)\\   
         \bm{r}_2 - \bm{r}_{i} & (i = 3,\dots, n)
     	\end{array}  
		\right.
			\label{eq:075}
\end{equation}
and taking into account that the Jacobian of this transformation is equal in modulus to $1$ because of the triangular form of the functional matrix, we obtain (omitting the primes)
\begin{equation}
\int_{x_2> 0, x_3,  \dots, x_i} x_2  {\cal U}^{(n)}_{0,2...n} d\bm{r}_2...d\bm{r}_n = \int_{x_2> 0, x_3>0, \dots, x_i>0} x_2  {\cal U}^{(n)}_{0,2...n} d\bm{r}_2...d\bm{r}_n
\label{eq:076}.
\end{equation}

Here we used the invariance of the Ursell factors under the transformation \(\ref{eq:075}\)\ (see the property \(\ref{eq:a022}\) for $i = $1, which is used in the transformation \(\ref{eq:a021}\)\).
 
The integral on the right-hand side of \(\ref{eq:076}\) is not an orthant moment because the variables $x_{i +1}, \dots, x_n$ may have arbitrary signs. But it is obviously the sum of orthant moments. That is,
\begin{equation}
\int_{x_2> 0, x_3,  \dots, x_i} x_2  {\cal U}^{(n)}_{0,2...n} d\bm{r}_2...d\bm{r}_n = \sum_{j=0}^{n-i} \binom{n-i}{j} [i+j,n]_1
\label{eq:077}.
\end{equation}

The second group of integrals on the right-hand side of \(\ref{eq:072}\) is divided into two parts $ x_2>0$ and $x_2<0$. With the first part, the problem reduces to the previous one \(only the index $i$ is increased by 1\).
 
The second part of integrals of this group \ ($ 0> x_2> x_3, \dots, x_ {i +1} $) is transformed by the change of variables
\begin{equation}
\bm{r}'_i =  
	 \left\{ 
			\begin{array}{ll} 
         - \bm{r}_{2} & (i = 2)\\   
         \bm{r}_{i} - \bm{r}_2 & (i = 3,\dots, n).
     	\end{array}  
		\right.
			\label{eq:078}
\end{equation}

Performing similar calculations and using \(\ref{eq:a027}\) for $l = 2$, we arrive at the expression
\begin{equation}
\int_{0> x_2> x_3,  \dots, x_{i+1}} x_2  {\cal U}^{(n)}_{0,2...n} d\bm{r}_2...d\bm{r}_n = - \sum_{j=0}^{n-i-1} \binom{n-i-1}{j} [j+2,n]_1
\label{eq:079}.
\end{equation}

Summation of \(\ref{eq:072}\) yields
\begin{equation}
I_n = (n-1)(1-\lambda)\sum_{k=0}^{n-2} \binom{n-2}{k}(\lambda^{n-k-2} - \lambda^{k+1}) [k+2,n]_1
\label{eq:080}.
\end{equation}

Multiplying \(\ref{eq:080}\) by $z^n$, we see that this expression is symmetric with respect to permutation of $z$ and $z'$
\begin{equation}
z^n I_n = (n-1)(z-z')\sum_{k=0}^{n-2} \binom{n-2}{k}(z^{k+1}z'^{n-k-2} - z'^{k+1}z^{n-k-2} ) [k+2,n]_1
\label{eq:081}.
\end{equation}

Collecting terms with equal powers of $\lambda$, we bring \(\ref{eq:080}\) to the form
\begin{multline}
I_n =(n-1)(1-\lambda)\Big \{ [n,n]_1 (1-\lambda^{n-1}) \\+  \sum_{k=1}^{n-2}\Big ( \binom{n-2}{k} [n-k,n]_1 -  \binom{n-2}{k-1} [k+1,n]_1\Big)\lambda^k \Big \}
\label{eq:082}.
\end{multline}

The expression \(\ref{eq:071}\) with $I_n$ in the form of \(\ref{eq:082}\) gives a variant of the surface cluster expansion. Other variants will be considered in the next section and appendix \ref{sec:b}.

\section{\label{sec:06}Surface quantities and the pair Ursell function}

Consider the first-order moment of ${\cal F}^{(2)}_{0,2}$ \(\ref{eq:020}\)
\begin{equation}
M^{(2)}_1(z,\lambda) = \int x_2 {\cal F}^{(2)}_{0,2} d\bm{r}_2
\label{eq:083},
\end{equation}
where the integration is performed over infinite space. Here ${\cal F}^{(2)}_{0,2}$ refers to the system in the step filed, and the $x_2$ axis is perpendicular to the field discontinuity surface on which the coordinate origin  is located.
 
Using \(\ref{eq:023}\), we obtain
\begin{equation}
M^{(2)}_1(z,\lambda) =  \theta_0 \sum_{n=2}^\infty \frac{z^n}{(n-2)!}J_n
\label{eq:084},
\end{equation}
where
\begin{equation}
J_n =  \int x_2 \Big [ \prod_{i = 2}^{n} \theta_i \Big ]  {\cal U}^{(n)}_{0,2...n} d\bm{r}_2...d\bm{r}_n.
\label{eq:085}
\end{equation}

We first calculate $\widetilde{J}_n$ - an analog of the $J_n$ \(\ref{eq:085}\) over the half-space $x _2> 0$,
\begin{equation}
\widetilde{J}_n =  \int\limits_{x_2>0} x_2 \Big [  \prod_{i = 2}^{n} \theta_i \Big ]  {\cal U}^{(n)}_{0,2...n} d\bm{r}_2...d\bm{r}_n
\label{eq:086}.
\end{equation}

Expressing the Boltzmann factors in terms of the characteristic functions \(\ref{eq:051}\) and \(\ref{eq:052}\)
\begin{equation}
\theta_i = \lambda \psi_i + \chi_i
\label{eq:087},
\end{equation}
we obviously obtain 
\begin{equation}
\widetilde{J}_n = \sum_{k=0}^{n-2} \binom{n-2}{k} \lambda^k [n-k,n]_1
\label{eq:088}.
\end{equation}

Returning to $J_n$, we get
\begin{equation}
{J}_n = \widetilde{J}_n(\lambda) - \lambda^{n-1} \widetilde{J}_n(\lambda^{-1}) 
\label{eq:089},
\end{equation}
where the second term on the right-hand side  \(the integral over the half-space $x_2 < 0$\) is also determined by substituting \(\ref{eq:087}\).
 
Substituting \(\ref{eq:088}\) into \(\ref{eq:089}\) and comparing the result with \(\ref{eq:082}\), we have
\begin{equation}
I_n = (1-\lambda)(n-1)J_n
\label{eq:090},
\end{equation}
or, in view of \(\ref{eq:071}\) and \(\ref{eq:084}\),
\begin{equation}
\beta z  \bigg ( \frac{\partial \nu}{\partial z} \bigg )_{ T } = (1-\lambda)\theta_0^{-1}M^{(2)}_1
\label{eq:091}.
\end{equation}

This is the relation of interest to us. It relates two fundamental quantities: the surface number density at the step boundary \(see below\) and the first-order moment of the pair Ursell function for this inhomogeneous medium.

Substituting \(\ref{eq:023}\) into \(\ref{eq:091}\) and integrating over the activity at a constant temperature, we return to the expression of $\nu $ as a series in $z$. Since the second-order Ursell function is determined by the number density and the pair distribution function \(\ref{eq:020}\) \(i.\,e., quantities subject to constant attention \), this form is of particular importance in comparison with other variants \(section \ref{sec:05}, appendix \ref{sec:b}\).

The factor $\theta_0^{-1}$ provides symmetry of the solution with respect to permutation of the solid and fluid regions \(with a simultaneous change of the signs of $u_0$ and $x$ and a change in the activity\), which must obviously be true for finite fields \(see also section \ref{sec:08}\).

Indeed, if the observation point is to the right of the step \(at the position $+0$\), then $\theta_0 = 1$ and we obtain
\begin{equation}
\beta z  \bigg ( \frac{\partial \nu}{\partial z} \bigg )_{ T } = (1-\lambda)M^{(2)}_1(z,\lambda)
\label{eq:092}.
\end{equation}

If the observation point is to the left of the step \(set to $ -0 $\), then $\theta_0 = \lambda $ and we arrive at the relation
\begin{equation}
\beta z  \bigg ( \frac{\partial \nu}{\partial z} \bigg )_{ T } = -\frac{1}{\lambda}(1-\lambda)M^{(2)}_1 = (1-\frac{1}{\lambda})M^{(2)}_1(\lambda z, \frac{1}{\lambda})
\label{eq:093},
\end{equation}
which is in agreement with \(\ref{eq:116}\).
 
The equation \(\ref{eq:091}\) may be viewed as a surface analog of the compressibility equation for a homogeneous medium
\begin{equation}
\beta z^2 \bigg ( \frac{\partial^2 P}{\partial z^2}  \bigg )_{ T } = M^{(2)}_0
\label{eq:094},
\end{equation}
where
\begin{equation}
M^{(2)}_0(z) =   \int {\cal F}^{(2)}_{0,2} d\bm{r}_2
\label{eq:095},
\end{equation}
is the zero-order moment of the pair Ursell function.
 
In view of \(\ref{eq:046}\), the equation \(\ref{eq:091}\) can be written as
\begin{equation}
\varrho_{s,n} = (\lambda - 1)\theta_0^{-1}M^{(2)}_1
\label{eq:096}.
\end{equation}

The volume analog of \(\ref{eq:096}\) is transformed to
\begin{equation}
\varrho(\varrho k_B T \varkappa_T -1) = M^{(2)}_0
\label{eq:097},
\end{equation}
where $\varkappa_T$ is the isothermal compressibility.
 
The expression \(\ref{eq:091}\) for $\lambda \rightarrow 0$ if the observation point is to the right of the boundary or for $\lambda \rightarrow \infty$ if the observation point is to the left \(low temperatures\) is fully consistent with the results of \cite{Zaskulnikov201102a} for a rigid impermeable wall. This becomes clear if we take into account the form of the series \(\ref{eq:023}\) for this case. Furthermore, this result also corresponds to the surface parameter of an open statistical ensemble because of the identity of the probabilistic and potential constraints \cite{zaskulnikov201004a}.

\section{\label{sec:07}High temperatures}

Consider the case of high temperatures \($\lambda \rightarrow 1$\). Let
\begin{equation}
\lambda = 1 + \varepsilon = 1 - \beta u_0 + \dots
\label{eq:098},
\end{equation}
then, from \(\ref{eq:089}\), we obtain
\begin{multline}
{J}_n = \widetilde{J}_n(1) + \varepsilon \widetilde{J}'_n(1) - \big[1 + (n-1)\varepsilon\big]\big[\widetilde{J}_n(1) - \varepsilon \widetilde{J}'_n(1)\big]+ \dots \\  = \varepsilon\big[2\widetilde{J}'_n(1) - (n-1)\widetilde{J}_n(1)\big] + \dots
\label{eq:099},
\end{multline}
where the prime denotes differentiation with respect to the argument.

From \(\ref{eq:a043}\) it follows immediately that
\begin{equation}
(n-2)\widetilde{J}_n(1) = 4 \widetilde{J}'_n (1)
\label{eq:100},
\end{equation}
and, thus, up to higher-order small quantities,
\begin{equation}
{J}_n = -\frac{\varepsilon}{2} n   \widetilde{J}_n (1)
\label{eq:101},
\end{equation}
or, in view of \(\ref{eq:090}\),
\begin{equation}
{I}_n = \frac{\varepsilon^2}{2} n(n-1)   \widetilde{J}_n (1)
\label{eq:102}.
\end{equation}

Substituting \(\ref{eq:102}\) into \(\ref{eq:071}\) and taking into account   \(\ref{eq:023}\) and \(\ref{eq:086}\), we obtain the main result:
\begin{equation}
\nu =\beta \frac{u_0^2}{2} \int\limits_{x_2 >0} x_2 {\cal F}^{(2)}_{0,2} d\bm{r}_2
\label{eq:103}.
\end{equation}

This expression relates the nonlinear surface coefficient in the limit of small step size and the second-order Ursell function of a \textit{homogeneous} medium.
 
We show that \(\ref{eq:103}\) is consistent with the results of \cite{Zaskulnikov201111a} for the potential of an arbitrary form.
 
We start from the expression given in \cite{Zaskulnikov201111a}, which in the case of a step potential takes the form
\begin{equation}
\bigg (\frac{\partial \nu}{\partial \mu}  \bigg )_{ T }=  (\beta u_0)^2 \Big [ \int \limits_{x_2>0} x_2 ~ {\cal F}^{(2)}_{0,2} d\bm{r}_2 + \int \limits_{x_2 > x_3} x_2 ~ {\cal F}^{(3)}_{0,2,3} d\bm{r}_2 d\bm{r}_3 \Big ] + \dots
\label{eq:104}
\end{equation}

Non-triviality is in the second integral on the right-hand side of \(\ref{eq:104}\), which includes the  Ursell function of the third rank. To bring it to the desired form, we  make use of the symmetry of this function with respect to coordinate inversion, permutations of particles, and translation. We consider two changes of variables with the transformation Jacobian of unit modulus. The first substitution, $\bm{r}'_2 = \bm{r}_2, \bm{r}'_3 = \bm{r}_2 - \bm{r}_3$, leads to the transformation
\begin{multline}
\int\limits_{x_2 > x_3} x_2 ~ {\cal F}^{(3)}_{0,2,3} d\bm{r}_2 d\bm{r}_3 = \int\limits_{x_3>0} x_2 ~ {\cal F}^{(3)}_{0,2,2-3} d\bm{r}_2 d\bm{r}_3 \\ = \int\limits_{x_3 > 0} x_2 ~ {\cal F}^{(3)}_{-2,0,-3} d\bm{r}_2 d\bm{r}_3 = \int\limits_{x_3 > 0} x_2 ~ {\cal F}^{(3)}_{2,0,3} d\bm{r}_2 d\bm{r}_3 = \int\limits_{x_3 > 0} x_2 ~ {\cal F}^{(3)}_{0,2,3} d\bm{r}_2 d\bm{r}_3
\label{eq:105}.
\end{multline}

The second change, $\bm{r}'_2 = \bm{r}_2 - \bm{r}_3, \bm{r}'_3 = -\bm{r}_3$, gives
\begin{multline}
\int\limits_{x_2 > x_3} x_2 ~ {\cal F}^{(3)}_{0,2,3} d\bm{r}_2 d\bm{r}_3  = \int\limits_{x_2 > 0} (x_2 - x_3) ~ {\cal F}^{(3)}_{0,2-3,-3} d\bm{r}_2 d\bm{r}_3  \\ =  \int\limits_{x_2 > 0} (x_2 - x_3) {\cal F}^{(3)}_{3,2,0} d\bm{r}_2 d\bm{r}_3 = \int\limits_{x_2 > 0} x_2 {\cal F}^{(3)}_{0,2,3} d\bm{r}_2 d\bm{r}_3 - \int\limits_{x_3 > 0} x_2 {\cal F}^{(3)}_{0,2,3} d\bm{r}_2 d\bm{r}_3 
\label{eq:106},
\end{multline}
where in the last integral, we made one more change $\bm{r}'_2 = \bm{r}_3, \bm{r}'_3 = \bm{r}_2$.

From \(\ref{eq:105}\) and \(\ref{eq:106}\), we obtain the relation of interest to us:
\begin{equation}
\int\limits_{x_2 > x_3} x_2 ~ {\cal F}^{(3)}_{0,2,3} d\bm{r}_2 d\bm{r}_3  = \frac{1}{2} \int\limits_{x_2 > 0} x_2 ~ {\cal F}^{(3)}_{0,2,3} d\bm{r}_2 d\bm{r}_3
\label{eq:107}.
\end{equation}

It has been shown previously \cite{zaskulnikov200911a} that
\begin{equation}
\int {\cal F}^{(3)}_{0,2,3} d\bm{r}_3  = z^3 \frac{\partial}{\partial z} \Big ( \frac{{\cal F}^{(2)}_{0,2}}{z^2} \Big )  
\label{eq:108}.
\end{equation}

Thus, substituting \(\ref{eq:107}\) into \(\ref{eq:104}\) and using \(\ref{eq:108}\), we obtain
\begin{equation}
\bigg (\frac{\partial \nu}{\partial \mu}  \bigg )_{ T } =  \frac{(\beta u_0)^2}{2} z \frac{\partial}{\partial z}\int \limits_{x_2>0} x_2 ~ {\cal F}^{(2)}_{0,2} d\bm{r}_2
\label{eq:109},
\end{equation}
which is seen to be consistent with \(\ref{eq:103}\).
 
In \cite{zaskulnikov201004a}, it is shown that the first-order moment of the pair Ursell function defines the surface part of the rms fluctuation in the number of particles \(at arbitrary temperatures\). Thus, at high temperatures, the nonlinear surface coefficient is uniquely related to this quantity.

\section{\label{sec:08}Symmetry of the problem}

In contrast to the adsorption problem with an impermeable wall, the absorption problem, in principle, is symmetric. By varying the parameters, we may consider a particular region of the fluid to be free or situated in the field. This issue in a general setting is considered in \cite{Zaskulnikov201111a}.

Making the substitution 
\begin{equation}
\begin{aligned}
& z^* = z'  \\
& z'^* = z  \\
& x^* = -x  \\
& u^*(x^*) = u(x) - u_0 \\
& u^*_0 = - u_0 
\end{aligned}
\label{eq:110}
\end{equation}
from \(\ref{eq:029}\) and \(\ref{eq:030}\) we make sure that
\begin{equation}
\begin{aligned}
& \widetilde{\theta}[u(x),u_0] = \widetilde{\theta}[u^*(x^*)-u^*_0,-u^*_0]= \widetilde{\varphi}[u^*(x^*),u^*_0]  \\
& \widetilde{\varphi}[u(x),u_0] = \widetilde{\varphi}[u^*(x^*)-u^*_0,-u^*_0] = \widetilde{\theta}[u^*(x^*),u^*_0].
\end{aligned}
\label{eq:111}
\end{equation}

Thus, one can see, for example, from \(\ref{eq:034}\) and \(\ref{eq:035}\), the coefficient $\nu$ is invariant under the transformation \(\ref{eq:110}\)
\begin{equation}
\nu(\widetilde{\theta},\widetilde{\varphi}|z,z') = \nu(\widetilde{\varphi}^*,\widetilde{\theta}^*|z'^*,z^*) =\nu(\widetilde{\theta}^*,\widetilde{\varphi}^*|z^*,z'^*) 
\label{eq:112},
\end{equation}
where we have introduced the notation
\begin{equation}
\begin{aligned}
& \widetilde{\theta}^* = \widetilde{\theta}[u^*(x^*),u^*_0]  \\
& \widetilde{\varphi}^* =  \widetilde{\varphi}[u^*(x^*),u^*_0], 
\end{aligned}
\label{eq:113}
\end{equation}
and the dependence of $\nu$ on $\widetilde{\theta}, \widetilde{\varphi}$ in \(\ref{eq:112}\) is functional.
 
In the derivation of \(\ref{eq:112}\), it is necessary to use the invariance of the Ursell factors under inversion \(\ref{eq:067}\).
 
In the step potential problem considered in this paper, the specific form of the potential in the transition region gives rise to the additional condition
\begin{equation}
\widetilde{\theta}(x) = \widetilde{\varphi}(-x)
\label{eq:114},
\end{equation}
which leads to a stronger consequence than \(\ref{eq:112}\): 
\begin{equation}
\nu(z,z') = \nu(z^*,z'^*) 
\label{eq:115}.
\end{equation}

This condition is equivalent to
\begin{equation}
\nu(z,\lambda) = \nu(z\lambda,\frac{1}{\lambda})
\label{eq:116}
\end{equation}
\(cf. \(\ref{eq:071}\)\) or
\begin{equation}
I_n(\lambda) = \lambda^n I_n(\frac{1}{\lambda})
\label{eq:117}
\end{equation}
\(cf. \(\ref{eq:080}\)\). \(As noted above, the equation \(\ref{eq:116}\) is consistent with the symmetry of the first-order moment of the Ursell function \(section \ref{sec:06}\).\)

In other words, the substitution \(\ref{eq:110}\) in this case reduces to the symmetric form
\begin{equation}
\begin{aligned}
& z^* = z'  \\
& z'^* = z  \\
& x^* = -x,  
\end{aligned}
\label{eq:118}
\end{equation}
or  the asymmetric form
\begin{equation}
\begin{aligned}
& z^* = \lambda z  \\
& \lambda^* = \frac{1}{\lambda}  \\
& x^* = -x.  
\end{aligned}
\label{eq:119}
\end{equation}

From \(\ref{eq:058}\) it follows that
\begin{equation}
\varrho(x',z,\lambda) = \varrho(-x',\lambda z,\frac{1}{\lambda})
\label{eq:120},
\end{equation}
and \(\ref{eq:057}\) implies that
\begin{equation}
\gamma(x',z,\lambda) = \gamma(-x',\lambda z,\frac{1}{\lambda})
\label{eq:121},
\end{equation} 
if we take into account that in the transformation \(\ref{eq:119}\),
\begin{equation}
x^*_{step} = - x_{step}
\label{eq:122}.
\end{equation}
 
The Henry constant  \(\ref{eq:048}\) satisfies the equation
\begin{equation}
K_H(x',\lambda) = \lambda K_H(-x',\frac{1}{\lambda})
\label{eq:123}
\end{equation}
in agreement with the  general result \cite{Zaskulnikov201111 a}.

\section{\label{sec:09}Generalized   contact   theorem}

We employ the orthant moments technique \(appendix \ref{sec:a}\) to generalize the contact theorem. Using this formalism, we first prove the usual contact theorem.
 
Consider a distribution function of the first rank in the field of a rigid impermeable wall. As follows from \(\ref{eq:022}\), it has the form
\begin{equation}
\varrho(\bm{r}_1) = z + \sum_{n=1}^\infty \frac{z^{n+1}}{n!} \int_{x_2>0, \dots, x_{n+1}>0} {\cal U}^{(n+1)}_{1,2...n+1} d\bm{r}_{2}...d\bm{r}_{n+1},
\label{eq:124}
\end{equation}
provided that the bounding field is in the range of negative $x$, and $x_1 > 0$.
 
Let $\bm{r} _1$ lie in the discontinuity plane of the potential; then in terms of the zero-order orthant moments \(\ref{eq:a009}\), this equation has the form
\begin{equation}
\varrho(+0) = z + \sum_{n=1}^\infty \frac{z^{n+1}}{n!} [n+1,n+1]_0
\label{eq:125}.
\end{equation}

Using the property \(\ref{eq:a015}\), we arrive at the series
\begin{equation}
\varrho(+0) = z + 2 \sum_{n=1}^\infty \frac{z^{n+1}}{(n+1)!} [n+1]_0
\label{eq:126}.
\end{equation}

Comparing this expression with \(\ref{eq:009}\), we see that the right-hand side of \(\ref{eq:126}\) is proportional to the pressure, and we obtain
\begin{equation}
\varrho(+0) = \beta P(z,T)
\label{eq:127},
\end{equation}
the well-known contact theorem \cite [p.\,166]{HansenMcDonald 2006}.
 
We now turn to the generalized contact theorem. In the expression \(\ref{eq:022}\) in this case, it is necessary to retain the Boltzmann factors. Using the expression \(\ref{eq:087}\) and passing to the orthant moments \(\ref{eq:a009}\) in \(\ref{eq:022}\), similarly to \(\ref{eq:125}\), we obtain
\begin{equation}
\frac{\varrho(\pm 0)}{\theta(\pm 0)} = z + \sum_{n=1}^\infty \frac{z^{n+1}}{n!} \sum_{m=0}^{n} \binom{n}{m} \lambda^m [m+1,n+1]_0
\label{eq:128},
\end{equation}
where the sign corresponds to the position of the observation point and where we have used the invariance of the Ursell functions with respect to particle permutations and the symmetry of the zero-order orthant moments \(\ref{eq:a012}\).
 
Here
\begin{equation}
\begin{aligned}
& \theta(+ 0) = 1  \\
& \theta(- 0) = \lambda 
\end{aligned}
\label{eq:129}
\end{equation}
and, therefore,
\begin{equation}
\frac{\varrho(- 0)}{\varrho(+ 0)} =  \lambda
\label{eq:139}. 
\end{equation}

Using \(\ref{eq:a011}\), we arrive at
\begin{equation}
\frac{\varrho(\pm 0)}{\theta(\pm 0)} = z + \sum_{n=1}^\infty \frac{z^{n+1}}{n!} [n+1,n+1]_0 \sum_{m=0}^{n} \lambda^m 
\label{eq:131}.
\end{equation}

Performing the inner summation and passing to integrals over the half-space by means of \(\ref{eq:a015}\), we obtain
\begin{equation}
\frac{\varrho(\pm 0)}{\theta(\pm 0)} = z + 2 \sum_{n=1}^\infty \frac{z^{n+1}}{(n+1)!} \frac{(1-\lambda^{n+1})}{(1-\lambda)} [n+1]_0 
\label{eq:132}.
\end{equation}

Now we can again use the expression for the pressure \(\ref{eq:009}\); then,
\begin{equation}
\frac{\varrho(\pm 0)}{\theta(\pm 0)} = z + \frac{1}{(1-\lambda)} \big [ \beta P(z) - z - \beta P(z \lambda) + z \lambda \big ]
\label{eq:133}
\end{equation}
or
\begin{equation}
\frac{\varrho(\pm 0)}{\theta(\pm 0)} =  \frac{\beta}{(1-\lambda)} \big[  P(z)  -  P(z \lambda) \big]
\label{eq:134}.
\end{equation}

This expression is the desired result - the generalized contact theorem.
 
In the limit $\lambda \rightarrow 0$ \(low temperatures\) or $\lambda \rightarrow \infty$ \(large negative potentials\), as one can easily see, the expression \(\ref{eq:134}\) transforms to the usual contact theorem \(\ref{eq:127}\).
 
For $\lambda \rightarrow 1$\(the case of a homogeneous medium or high temperatures\) \(\ref{eq:134}\) becomes the differential equation \(\ref{eq:011}\).
 
It is easy to verify that the solution has the required symmetry property \(\ref{eq:120}\)
\begin{equation}
\varrho(\pm 0, z, \lambda) =  \varrho(\mp 0, z \lambda, \frac{1}{\lambda})
\label{eq:135},
\end{equation}
if we consider that
\begin{equation}
\frac{\theta( - 0, \displaystyle \frac{1}{\lambda})}{\theta( + 0, \lambda)} =  \frac{1}{\lambda}
\label{eq:136}.
\end{equation}

Equation \(\ref{eq:134}\) under these assumptions can be written as
\begin{equation}
\varrho(+0) - \varrho(-0) = \beta \big[  P(+\infty)  -  P(-\infty) \big]
\label{eq:137},
\end{equation}
where the pressure and number density are considered in the half-space of the corresponding sign.

\section{\label{sec:10}Summary}

\begin{enumerate}
	\item For the step potential model, expressions were obtained for the specific surface $\Omega$-potential $\gamma$\(\ref{eq:057}\), the surface nonlinear coefficient $\nu$\(\ref{eq:055}\), and the surface number density \(\ref{eq:058}\).		
	\item The zero adsorption surface in this case coincides with the discontinuity plane of the potential \(\ref{eq:056}\).	
	\item The Henry adsorption constant depends on the position of the separating boundary and is given by \(\ref{eq:059}\).		
	\item The adsorption and absorption Henry constants are anticorrelated and their relationship is given by \(\ref{eq:060}\).  	
	\item Expressions were obtained for the nonlinear surface coefficient $\nu $ as a series in powers of the activity - an analog of the group series for pressure - \(\ref{eq:071}\), \(\ref{eq:082}\), section \ref{sec:06} and appendix \ref{sec:b}. The zero-order moments of the Ursell factors are replaced by sums of the first-order moments over orthants and sectors \(appendix \ref{sec:a}\). 		
	\item The invariant part of the surface number density at a stepped boundary is determined by the first-order moment of the pair Ursell function for an \textit{inhomogeneous} medium \(section \ref{sec:06}\).		
	\item At high temperatures \(small step size\), the nonlinear surface coefficient $\nu$ is determined by the first-order moment of the pair Ursell function of a \textit{homogeneous} medium over a half-space \(section \ref{sec:07}\). Thus, it is related to the surface part of the standard deviation of the number density.  		
	\item The step potential problem has the additional \(compared to the general problem statement\) property of symmetry with respect to permutation of the field and free fluid regions \(section \ref{sec:08}\).    	
	\item The well-known contact theorem is extended to the case of a step potential of arbitrary magnitude \(section \ref{sec:09}\).			
\end{enumerate}

\appendix
\begin{appendices}
\numberwithin{equation}{section}

\section{\label{sec:a}Sectorial and orthant moments}

In this section, we consider the zero- and first-order moments of functions that have certain symmetry properties. Regardless of the specific form of these functions, it is possible to derive general relations which lead to a number of important results. As such functions we will use the Ursell factors.
 
These moments will be taken over some regions: sectors and orthants \(see below\). The values of one integration variable \($x$\) are restricted by certain conditions, and the other variables \($y, z$\) are independent and take values corresponding to the infinite space.
 
Sectorial moments play an auxiliary role and are required mainly to derive relations for the orthant moments. Ultimately, we are interested in integrals over whole space and half-space.

\subsection{\label{subsec:a01}Sectors and orthants}

The terms ``sector'' and ``orthant'' are used to refer to a consideration of a geometric picture in the space with coordinates ${x_2, x_3, \dots, x_n}$.

By a sector we mean a region
\begin{equation}
x_{\alpha_1} > x_{\alpha_2} > \dots > x_{\alpha_{j-1}} > 0 > x_{\alpha_{j+1}}> \dots > x_{\alpha_n}
\label{eq:a001},
\end{equation}
where $\alpha_i$ correspond to different particles and take values from $2$ to $n$.
 
\begin{figure}[htbp] 
	\center{
	  \includegraphics{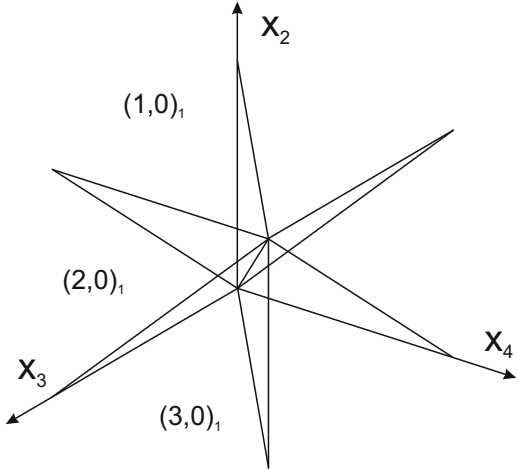}}
	  \caption{Some sectors and the corresponding sectorial moments \(of the first order\) for $n=4$. The moment-generating variable is $x_2$.}
   \label{fig:01}
\end{figure}

The position of the zero in the chain \(\ref{eq:a001}\) - $j$ - can take values  $1,\dots, n$ for the zero-order moments and $2,\dots, n$ for the first-order moments. It plays an important role and, in particular, determines the shape and size of the sector. The simplest example is given in Fig. \ref{fig:01}.

The orthant is the region
\begin{equation}
x_{\alpha_1} > 0, ~ x_{\alpha_2} > 0, \dots ,  ~x_{\alpha_{j-1}} > 0, ~~x_{\alpha_{j+1}}<0,  \dots , ~x_{\alpha_n}<0
\label{eq:a002}.
\end{equation}

The total number of orthants is $2^{n-1}$ for the zero-order moments and $2^{n-2}$ for the first-order moments.
 
Obviously, an orthant is the union of a certain number of sectors. The number of sectors in a given orthant is defined by permutations of particles within the positive and negative groups and it is equal to 
\begin{equation}
(j-1)!(n-j)!
\label{eq:a003}.
\end{equation}

We now show that, due to the translational, permutation, and inversion symmetry of the Ursell factors, the sectorial and orthant moments satisfy some relations.\footnote{Of course, the relations obtained below are valid not only for the Ursell factors but also for arbitrary functions with the same symmetry properties. In particular, they are valid for the products of the Mayer functions  \scriptsize(\footnotesize\ref{eq:031}\scriptsize) \footnotesize and for Ursell \textit{functions} of type \scriptsize(\footnotesize\ref{eq:020}\scriptsize)\footnotesize.}

\subsection{\label{subsec:a02}Zero-order moments}

\textbf{Sectorial moments.} In the zero-order case, the sectorial moments are defined by the equality
\begin{equation}
(i)_0 = \int_{x_2 > x_3 > \dots > x_{i+1} > 0 > x_{i+2} > \dots > x_n}  ~ {\cal U}^{(n)}_{0,2...n} d\bm{r}_2...d\bm{r}_n
\label{eq:a004},
\end{equation}
where $i$ is the number of positive coordinates in the chain \(\ref{eq:a001}\), \($i=0, \dots, n-1$\). We note that the $x$ coordinates of \textit{all} particles should have a pronounced relation to zero.

The change of variables
\begin{equation}
\bm{r}'_k =  
	 \left\{ 
			\begin{array}{ll} 
         - \bm{r}_2 & (k = 2)\\   
         \bm{r}_k - \bm{r}_{2} & (k = 3,\dots, n)
     	\end{array}  
		\right.
			\label{eq:a005}
\end{equation}
transforms the right-hand side of \(\ref{eq:a004}\) as follows.

The Ursell factor:
\begin{multline}
 {\cal U}^{(n)}_{0,2,3,\dots,n} = {\cal U}^{(n)}(0,\bm{r}_2,\bm{r}_3,\dots,\bm{r}_n) = {\cal U}^{(n)}(0,-\bm{r}'_2,\bm{r}'_3-\bm{r}'_{2},\dots,\bm{r}'_n-\bm{r}'_{2}) \\ = {\cal U}^{(n)}(\bm{r}'_{2},0,\bm{r}'_3,\dots,\bm{r}'_n) = {\cal U}^{(n)}(0,\bm{r}'_2,\bm{r}'_3,\dots,\bm{r}'_n)
\label{eq:a006},
\end{multline}
where in the penultimate equality we used the invariance of ${\cal U}^{(n)}$ under translation, and in the last equality, its invariance under particle permutations.

Sector \(region of integration\):
\begin{multline}
\{x_2 > x_3 > \dots >  x_{i+1} > 0 > x_{i+2} > \dots > x_n\} \\ = \{-x'_2 > x'_3 - x'_2 > \dots >  x'_{i+1} - x'_2 > 0 > x'_{i+2} - x'_2 > \dots > x'_n - x'_2\} \\ =  \{0 > x'_3 > \dots >  x'_{i+1} > x'_2 > x'_{i+2} > \dots > x'_n\}
\label{eq:a007}.
\end{multline}

Given that the Jacobian of the transformation \(\ref{eq:a005}\) is equal in modulus to $1$, we see that the zero-order sectorial moments are independent of the index $i$. Thus, we conclude that
\begin{equation}
(i)_0 = (0)_0, ~~i = 1,\dots, n-1.
\label{eq:a008}
\end{equation}

\vspace{10pt}

\textbf{Orthant moments.} In the zero-order case, the orthant moments are defined by
\begin{equation}
[m,n]_0 = \int_{x_2 > 0, \dots, x_m>0,  x_{m+1} < 0, \dots, x_n<0} ~ {\cal U}^{(n)}_{0,2...n} d\bm{r}_2...d\bm{r}_n
\label{eq:a009},
\end{equation}
where the index $m$ is equal to the number of positive coordinates plus one and $n$ is obviously equal to the total number of coordinates \(number of nonzero coordinates plus one\), so that $m= 1, \dots, n$.
 
Obviously, an arbitrary orthant moment is equal to the sum of the corresponding sectorial moments:
\begin{equation}
[m,n]_0 = (m-1)!(n-m)! (0)_0
\label{eq:a010},
\end{equation}
which reflects the partition of the orthant into sectors \(\ref{eq:a003}\). This equality allows any moment to be expressed in terms of the higher-order moment
\begin{equation}
[m,n]_0 = \binom{n-1}{m-1}^{-1}[n,n]_0
\label{eq:a011}.
\end{equation}

Note  that \(\ref{eq:a011}\) implies the symmetry property
\begin{equation}
[m,n]_0 = [n-m+1,n]_0
\label{eq:a012}.
\end{equation}

For the integral over the half-space
\begin{equation}
[n]_0 = \int\limits_{x_2 > 0}^{} ~ {\cal U}^{(n)}_{0,2...n} d\bm{r}_2...d\bm{r}_n 
\label{eq:a013},
\end{equation}
we obviously have
\begin{equation}
[n]_0 = \sum_{m=2}^{n} \binom{n-2}{m-2} [m,n]_0
\label{eq:a014}.
\end{equation}

From \(\ref{eq:a011}\), we obtain the relation
\begin{equation}
[n]_0 = \sum_{m=2}^{n} \frac{(m-1)}{(n-1)} [n,n]_0 = \frac{n}{2} [n,n]_0
\label{eq:a015},
\end{equation}
which, together with (\ref{eq:a011}), will be used  in the derivation of the contact theorems.

\subsection{\label{subsec:a03}First-order moments}

\textbf{Sectorial moments.} In the first-order case, they are defined by
\begin{equation}
(i,n-j)_1 = \int_{x_2 > \dots > x_j>0>x_{j+1}>\dots > x_n} x_{i+1} ~ {\cal U}^{(n)}_{0,2...n} d\bm{r}_2...d\bm{r}_n
\label{eq:a016},
\end{equation}
where $i$ is the position of the moment-generating variable \($i=1, \dots, n-1$\) in the chain of inequalities \(\ref{eq:a001}\) and $j$ is the position of the zero. Moreover, we assume that always $i <j$, i.\,e., the zero is always more distant in the chain than the moment-generating variable and that the latter is always positive. Thus, $j=i+1, \dots, n$. \(The index $n$ will always denote the rank of the Ursell factor.\) Obviously, the second index, $n-j$, is equal to the number of negative coordinates.

It follows that for an arbitrary sectorial moment of the first order
\begin{equation}
(m,k)_1
\label{eq:a017}
\end{equation}
the sum of the indices always satisfies the relation
\begin{equation}
m+k \leq n-1
\label{eq:a018}.
\end{equation}

Integrals of the type of \(\ref{eq:a016}\), where $x_{i +1} < 0$, are reduced to sectorial moments by the inversion \(\ref{eq:067}\).
 
Examples of the sectorial moments are
\begin{equation}
(1,0)_1 = \int_{x_2 > \dots > x_n>0} x_2 ~ {\cal U}^{(n)}_{0,2...n} d\bm{r}_2\dots d\bm{r}_n
\label{eq:a019}
\end{equation}
or
\begin{equation}
(2,n-3)_1 = \int_{x_3 > x_2 >0>  \dots > x_n} x_2 ~ {\cal U}^{(n)}_{0,2...n} d\bm{r}_2\dots d\bm{r}_n
\label{eq:a020}.
\end{equation}

The change of variables
\begin{equation}
\bm{r}'_k =  
	 \left\{ 
			\begin{array}{ll} 
         \bm{r}_{i+1} & (k = i + 1)\\   
         \bm{r}_{i+1} - \bm{r}_k & (k = 2,\dots, i, i+2,\dots,n)
     	\end{array}  
		\right.
			\label{eq:a021}
\end{equation}
transforms the right-hand side of \(\ref{eq:a016}\) as follows.
 
The Ursell factor:
\begin{multline}
{\cal U}^{(n)}_{0,2,\dots,i+1,\dots,n} = {\cal U}^{(n)}(0,\bm{r}_2,\dots,\bm{r}_{i+1},\dots,\bm{r}_n) \\  = {\cal U}^{(n)}(0,-\bm{r}'_2+\bm{r}'_{i+1},\dots,\bm{r}'_{i+1}, \dots,-\bm{r}'_n+\bm{r}'_{i+1})= {\cal U}^{(n)}(-\bm{r}_{i+1},-\bm{r}'_2,\dots,0,\dots,-\bm{r}'_n)\\ = {\cal U}^{(n)}(\bm{r}_{i+1},\bm{r}'_2,\dots,0,\dots,\bm{r}'_n) = {\cal U}^{(n)}(0,\bm{r}'_2,\dots,\bm{r}'_{i+1},\dots,\bm{r}'_n)
\label{eq:a022},
\end{multline}
where we have used the invariance of ${\cal U}^{(n)}$ with respect to translation, inversion, and permutations of particles.
 
The sector \(region of integration\):
\begin{multline}
\{x_2 > \dots > x_i >  x_{i+1}>  x_{i+2}> \dots > x_j>0>x_{j+1}> \dots > x_n\}  = \{-x'_2+x'_{i+1} > \\ \dots >   -x'_i+ x'_{i+1} > x'_{i+1}>  -x'_{i+2}+ x'_{i+1}> \dots > -x'_j +x'_{i+1}>0>-x'_{j+1}+x'_{i+1} > \dots \\ >  -x'_n+x'_{i+1}\} =  \{-x'_2 > \dots > -x'_i >  0 >  -x'_{i+2} > \dots > -x'_j>-x'_{i+1}>-x'_{j+1} > \\ \dots > -x'_n\}  =  \{x'_n > \dots > x'_{j+1} > x'_{i+1} > x'_j > \dots > x'_{i+2} >  0> x'_i > \dots > x'_2\}
\label{eq:a023}.
\end{multline}

Given that the Jacobian of the transformation \(\ref{eq:a021}\) is equal in modulus to $1$, we obtain  the relation 
\begin{equation}
(i,n-j)_1 = (n-j+1,i-1)_1
\label{eq:a024},
\end{equation}
where $i=1, \dots, n-1; ~~j=i+1, \dots, n $  or
\begin{equation}
(m,k)_1 = (k+1,m-1)_1
\label{eq:a025},
\end{equation}
where $m=1, \dots, n-1; ~~k=0, \dots, n-m-1$.
 
Equation \(\ref{eq:a025}\) is the first basic relation for the sectorial moments of the first order.
 
The change of variables
\begin{equation}
\bm{r}'_k =  
	 \left\{ 
			\begin{array}{ll} 
         -\bm{r}_{l} & (k = l)\\   
         \bm{r}_k - \bm{r}_l & (k = 2,\dots, l-1, l+1,\dots,n),
     	\end{array}  
		\right.
			\label{eq:a026}
\end{equation}
where $l$ is a fixed number in the range of $j +1, \dots, n$ yields a different transformation of the right-hand side of \(\ref{eq:a016}\).
 
The Ursell factor remains invariant
\begin{multline}
{\cal U}^{(n)}_{0,2,\dots,l,\dots,n} = {\cal U}^{(n)}(0,\bm{r}_2,\dots,\bm{r}_{l},\dots,\bm{r}_n) \\ = {\cal U}^{(n)}(0,\bm{r}'_2-\bm{r}'_{l},\dots,-\bm{r}'_{l},\dots,\bm{r}'_n-\bm{r}'_{l})= {\cal U}^{(n)}(\bm{r}'_{l},\bm{r}'_2,\dots,0,\dots,\bm{r}'_n) \\ = {\cal U}^{(n)}(0,\bm{r}'_2,\dots,\bm{r}'_{l},\dots,\bm{r}'_n)
\label{eq:a027},
\end{multline}
and the integration sector transforms as
\begin{multline}
\{x_2 > \dots >  x_{i+1}>  \dots > x_j>0>x_{j+1} > \dots > x_l> \dots> x_n\} = \{x'_2-x'_{l} > \dots \\  > x'_{i+1} -x'_l> \dots > x'_j -x'_{l}>0>x'_{j+1}-x'_{l} > \dots >-x'_l> \dots > x'_n-x'_{l}\} \\ =  \{x'_2 > \dots > x'_{i+1}> \dots > x'_j>x'_{l}>x'_{j+1} > \dots > x'_{l-1}>0>x'_{l+1} > \dots > x'_n\}  
\label{eq:a028}.
\end{multline}

In addition, in this case we obtain the transformation of the integrand factor
\begin{equation}
x_{i+1} = x'_{i+1}- x'_l
\label{eq:a029}
\end{equation}
and the integral on the right-hand side of \(\ref{eq:a016}\) splits into two.
 
Given that the Jacobian of the transformation \(\ref{eq:a026}\) is again equal in modulus to $1$, we obtain the second relation
\begin{equation}
(i,n-j)_1 = (i,n-l)_1 - (j,n-l)_1
\label{eq:a030},
\end{equation}
where $i=1, \dots, n-2; ~~j=i+1, \dots, n-1; ~~ l=j+1, \dots,n$. 
 
Note that the index $l$ is free, for given $i, j$, we have a whole class of relations \(\ref{eq:a030}\) corresponding to different $l$. For example, for $l = n$, we obtain
\begin{equation}
(i,n-j)_1 = (i,0)_1 - (j,0)_1
\label{eq:a031},
\end{equation}
$i=1, \dots, n-2; ~~j=i+1, \dots, n-1$.
 
By sequential decomposition of \(\ref{eq:a030}\), it is easy to show that all these representations are equivalent and we can use any of them, for example, \(\ref{eq:a031}\).
 
Thus, \(\ref{eq:a031}\) is the second basic equation for the first-order sectorial moments.
 
Substituting $i = 1$ and $j = n - i$ in \(\ref{eq:a031}\) and using \(\ref{eq:a025}\), we obtain
\begin{equation}
(n-i,0)_1 + (i+1,0)_1 = (1,0)_1
\label{eq:a032},
\end{equation}
where $i = 1, 2, \dots, n-2$, which allows eliminating the higher-order terms $(i,0)_1$ from the equations. Summing yields
\begin{equation}
\sum_{i=m+1}^{n-m}(i,0)_1 = (\frac{n}{2} - m)(1,0)_1
\label{eq:a033}
\end{equation}
and, in particular, the total sum is 
\begin{equation}
\sum_{i=1}^{n-1}(i,0)_1 = \frac{n}{2}(1,0)_1
\label{eq:a034}.
\end{equation}

\vspace{10pt}
\textbf{Orthant moments.} In the first-order case, the orthant moments are defined by the equality
\begin{equation}
[m,n]_1 = \int_{x_2 > 0, \dots, x_m>0,  x_{m+1} < 0, \dots, x_n<0} x_2 ~ {\cal U}^{(n)}_{0,2...n} d\bm{r}_2...d\bm{r}_n
\label{eq:a035},
\end{equation}
where the index $m$ is equal to the number of positive coordinates plus one and $n$ is obviously equal to the total number of coordinates \(the number of nonzero coordinates plus one\). We will consider only the case where the moment-generating variable is greater than zero, so that $m =2, \dots, n$. In the case $x_2 < 0$, we can use the inversion transformation \(\ref{eq:067}\).
 
For example, the orthant moment
\begin{equation}
[n,n]_1 = \int_{x_2 > 0, \dots, x_n>0} x_2 ~ {\cal U}^{(n)}_{0,2...n} d\bm{r}_2...d\bm{r}_n
\label{eq:a036}
\end{equation}
specifies the first-order moment of the Ursell factor over the orthant where all $x_i$ are positive.
 
Obviously, the orthant and sectorial moments of the first order are linked by the relation
\begin{equation}
[m,n]_1 = (m-2)!(n-m)! \sum_{k=1}^{m-1} (k,n-m)_1
\label{eq:a037},
\end{equation}
which reflects a simple geometric partition of orthants into sectors.

Expanding the moment on the right-hand side of \(\ref{eq:a037}\) with the use of \(\ref{eq:a031}\), and performing the summation, we obtain
\begin{equation}
[m,n]_1 =  (m-2)!(n-m)! \sum_{k=1}^{m-1} (k,0)_1  -  (m-1)!(n-m)!  (m,0)_1
\label{eq:a038},
\end{equation}
where $m = 2, 3, \dots, n-1$.
 
For $m = n$, we directly use \(\ref{eq:a037}\)
\begin{equation}
[n,n]_1 = (n-2)! \sum_{k=1}^{n-1} (k,0)_1
\label{eq:a039}.
\end{equation}

With the use of \(\ref{eq:a034}\), we make a further simplification
\begin{equation}
[n,n]_1 = \frac{n(n-2)!}{2}(1,0)_1
\label{eq:a040}.
\end{equation}

Equation \(\ref{eq:a038}\) defines an orthant moment in terms of sectorial moments. The inverse relation has the form
\begin{equation}
(m,0)_1 = \frac{[n-m+1,n]_1}{n(m-1)!(n-m-1)!} - \frac{[m,n]_1}{n(m-2)!(n-m)!}+\frac{[n,n]_1}{n(n-2)!}
\label{eq:a041},
\end{equation}
where $k=1, \dots,n -1$. The validity of \(\ref{eq:a041}\) is easily proved by using the direct substitution of \(\ref{eq:a038}\) and the relation \(\ref{eq:a040}\). Thus, the sectorial and orthant moments of the first order are rigidly connected.
 
By using \(\ref{eq:a038}\), \(\ref{eq:a040}\), and \(\ref{eq:a032}\) and by direct substitution, we can verify the validity of the equality
\begin{equation}
\sum_{m=2}^{n} \binom{n-2}{m-2} (3n-4m+2) [m,n]_1 = 0
\label{eq:a042},
\end{equation}
which will be employed to obtain high-temperature expansions. It will be convenient to bring to the form
\begin{equation}
\sum_{m=0}^{n-2} \binom{n-2}{m}  [n-m,n]_1 = 4 \sum_{m=1}^{n-2} \binom{n-3}{m-1}  [n-m,n]_1
\label{eq:a043},
\end{equation}
by replacing $m' = n - m$.

\section{\label{sec:b}Surface cluster expansion - third variant}

The orthant moments are included in the surface cluster expansion for the step potential \(\ref{eq:071}\), \(\ref{eq:082}\). Using the sorting identity \(appendix \ref{sec:c}\), we can transfer the consideration to the level of sectorial moments.
 
This identity \(\ref{eq:c006}\) is applicable to the right-hand side of \(\ref{eq:066}\). Using it, we obtain
\begin{equation}
{f}^{(n-1)}_{2...n} = - (1-\lambda)\sum_{i=1}^{n} \lambda^{i-1} x_{\beta_i}
\label{eq:b001},
\end{equation}
where the chain of elements is constructed so that 
\begin{equation}
x_{\beta_1} < x_{\beta_2} < \dots < x_{\beta_n}
\label{eq:b002}.
\end{equation}

Naturally, the first element \(zero\) can be in an arbitrary position. Its position is a key factor.
 
Since, as already mentioned, the Ursell factors are invariant under the inversion operation \(\ref{eq:067}\), we can assume that \ref{eq:069} holds, or

\begin{equation}
{f}^{(n-1)}_{2...n} =  (1-\lambda)\sum_{i=1}^{n} \lambda^{i-1} x_{\beta_i}
\label{eq:b003},
\end{equation}
and
\begin{equation}
x_{\beta_1} > x_{\beta_2} > \dots > x_{\beta_n}
\label{eq:b004}.
\end{equation}

Substituting the expression \(\ref{eq:b003}\) into \(\ref{eq:061}\), we obtain  the quantity $\nu$ of interest to us. In this case, since the integration is performed over the coordinates of each particle in an infinite space, it is necessary to consider all possible permutations in the chains \(\ref{eq:b004}\), bearing in mind that the zero can be in an arbitrary position.
 
The importance of the position of the zero is due to the fact that it actually separates one chain into two independent chains.
 
Obviously, for a given position of the zero, we have $(n-1)!$ identical integrals.
 
Thus, \(\ref{eq:061}\) becomes
\begin{equation}
\beta  \nu (z,\lambda) = (1 - \lambda) \sum_{n=2}^\infty \frac{z^n}{n} \sum_{i=1}^{n} \lambda^{i-1}\sum_{j=1}^{n} \int_{x_3 > \dots >0>\dots >x_2 >\dots >x_n} x_{2} ~ {\cal U}^{(n)}_{0,2...n} d\bm{r}_2...d\bm{r}_n
\label{eq:b005}.
\end{equation}

The second sum on the right-hand side of \(\ref{eq:b005}\) is over the position of the integration variable in the chain of inequalities, and the third over the position of the zero, which vary from the first position to the last. The diagonal elements  \(in the case of coincidence of the positions of the zero and  variable $x_2$\) are obviously zero.
 
The integrals on the right-hand side of \(\ref{eq:b005}\), up to sign, are sectorial moment of the first order \(appendix \ref{subsec:a03}\). For $ i<j$, we substitute \(\ref{eq:a016}\) into \(\ref{eq:b005}\), and for  $i>j $ we apply the inversion \(\ref{eq:067}\). Then, because in the transformation \(\ref{eq:067}\), $j'=n-j+1; i' = n-i +1$, we get for $i> j$ that the integral in \(\ref{eq:b005}\) takes the form
\begin{equation}
-(n-i+1,j-1)_1 = -(j,n-i)_1
\label{eq:b006},
\end{equation}
where the minus is due to the change in the sign of $x_2$. Here we have used \(\ref{eq:a025}\).
 
Thus, from \(\ref{eq:b005}\), we obtain
\begin{equation}
\beta  \nu (z,\lambda) = (1 - \lambda) \sum_{n=2}^\infty \frac{z^n}{n} \sum_{i=1}^{n} \lambda^{i-1}\sum_{j=1}^{n} C_{ij}
\label{eq:b007},
\end{equation}
where
\begin{equation}
C_{ij} = 
	 \left\{ 
			\begin{array}{rl} 
         (i,n-j)_1& (i<j)\\  
         0& (i=j) \\
         -(j,n-i)_1& (i>j).         
     	\end{array}  
		\right.
			\label{eq:b008}
\end{equation}

Expanding the moments in \(\ref{eq:b008}\) with the use of \(\ref{eq:a031}\) and taking into account that $(k,n-n)_1 = (k,0)_1$, we have
\begin{equation}
C_{ij} = 
	 \left\{ 
			\begin{array}{rl} 
				 (i,0)_1& (j=n)\\
         (i,0)_1 - (j,0)_1& (i,j\neq n)\\  
         -(j,0)_1& (i=n).
     	\end{array}  
		\right.
			\label{eq:b009}
\end{equation}

Summing yields
\begin{equation}
\beta  \nu (z,\lambda) =  \sum_{n=2}^\infty z^n \Big [ (1 - \lambda) \sum_{i=1}^{n-1} \lambda^{i-1}(i,0)_1- (1 - \lambda^n)\frac{(1,0)_1}{2}\Big ]
\label{eq:b010},
\end{equation}
where we have used \(\ref{eq:a034}\).
 
Taking out the factor $(1-\lambda)$  from $I_n$ \(\ref{eq:070}\), we obtain
\begin{equation}
I_n = (1-\lambda)n!\Big \{ \sum_{i=0}^{n-2} \lambda^i \Big [ (i+1,0)_1 - \frac{(1,0)_1}{2} \Big ] - \lambda^{n-1}\frac{(1,0)_1}{2}\Big\}
\label{eq:b011}
\end{equation}
or, placing $(1-\lambda) $ under the summation sign,
\begin{equation}
I_n = n! \Big \{ \frac{(1,0)_1}{2}+ \sum_{i=1}^{n-2} \lambda^i \Big [ (i+1,0)_1 - (i,0)_1\Big ] - \lambda^{n-1}(n-1,0)_1+\lambda^{n}\frac{(1,0)_1}{2}\Big \}
\label{eq:b012}.
\end{equation}

Symmetry of this expression in the form \(\ref{eq:117}\), as is easily verified, is provided by the relations \(\ref{eq:a032}\).
 
The expression \(\ref{eq:b011}\) is equivalent to \(\ref{eq:082}\), as is easily seen by equating the coefficients of like powers of $\lambda$. Their coincidence is determined by the satisfaction of the relation \(\ref{eq:a041}\).

\section{\label{sec:c}Sorting  identity}

Consider the set of $n$ elements to be compared \(ordered\)
\begin{equation}
 \{x_1, x_2, \dots, x_n  \}
\label{eq:c001}.
\end{equation}

Obviously, without loss of generality, we can assume that
\begin{equation}
 x_1< x_2< \dots < x_n 
\label{eq:c002}.
\end{equation}

We consider the structure
\begin{equation}
f = \sum_{k=1}^n (1-\lambda)^{k-1} \lambda^{n-k+1} \sum_{samp} \min ( x_{\alpha_1}, x_{\alpha_2}, \dots, x_{\alpha_k} )
\label{eq:c003},
\end{equation}
where the inner sum is over all possible samples of $k$ elements of $n$, and perform its calculation.

Taking into account \(\ref{eq:c002}\), we have the expression
\begin{equation}
f = \sum_{k=1}^n (1-\lambda)^{k-1} \lambda^{n-k+1} \sum_{l=1}^{n-k+1} \binom{n-l}{k-1} x_l
\label{eq:c004}
\end{equation}
and changing the order of summation, we obtain
\begin{equation}
f = \sum_{l=1}^n  x_l \sum_{k=1}^{n-l+1} \binom{n-l}{k-1} (1-\lambda)^{k-1} \lambda^{n-k+1} = \sum_{l=1}^n \lambda^{l} x_l
\label{eq:c005},
\end{equation} 
or, ultimately,
\begin{equation}
\sum_{l=1}^n \lambda^{l} x_l = \sum_{k=1}^n (1-\lambda)^{k-1} \lambda^{n-k+1} \sum_{samp} \min ( x_{\alpha_1}, x_{\alpha_2}, \dots, x_{\alpha_k} ) 
\label{eq:c006}
\end{equation}
if the condition \(\ref{eq:c002}\) is satisfied.
 
This is the desired identity, which may be called the sorting identity, because given the right-hand side of the equality, its left-hand side is automatically arranged in ascending order of $x_l$.
 
It generalizes the well-known maximum-minimums identity \cite{Ross2010}, which has the form
\begin{equation}
\max (x_1, \ldots , x_n) = \sum_{k=1}^n (-1)^{k-1} \sum_{samp} \min (x_{\alpha_1}, x_{\alpha_2}, \dots, x_{\alpha_k}  )
\label{eq:c007}.
\end{equation}

Indeed, in the limit $\lambda\rightarrow\infty$, the equality \(\ref{eq:c006}\) turns, obviously, in \(\ref{eq:c007}\).
 
As $\lambda\rightarrow 0$ or $\lambda\rightarrow 1$, (\ref{eq:c006}\) becomes trivial relations.
 
Using the substitution $x'_i = -x_i$, we obtain the second form of the sorting identity
\begin{equation}
\begin{gathered}
\sum_{l=1}^n \lambda^{l} x_l = \sum_{k=1}^n (1-\lambda)^{k-1} \lambda^{n-k+1} \sum_{samp} \max ( x_{\alpha_1}, x_{\alpha_2}, \dots, x_{\alpha_k} );    \\
 \text{provided that:} ~  x_1> x_2> \dots > x_n   ,
\label{eq:c008}
\end{gathered}
\end{equation}
which can be called the inverse-sorting identity corresponding to the second equivalent form of the maximum-minimums identity
\begin{equation}
\min(x_1, \ldots , x_n) = \sum_{k=1}^n (-1)^{k-1} \sum_{samp} \max (x_{\alpha_1}, x_{\alpha_2}, \dots, x_{\alpha_k}  )
\label{eq:c009}.
\end{equation}

\end{appendices}

\tiny 
\raggedleft
VZ, 30.05.2012, v045.

\end{document}